\documentclass[conference]{csce}

\usepackage[hmargin=.75in,vmargin=1in]{geometry}
\usepackage[american]{babel}
\usepackage[T1]{fontenc}
\usepackage{times}
\usepackage{caption}

\usepackage{textcomp}
\usepackage{epsfig,graphicx}
\usepackage{xcolor}
\usepackage{amsfonts,amsmath,amssymb}
\usepackage{fixltx2e} 
\usepackage{booktabs}

\title{\bf Machine Learning for Predictive Analytics of Compute Cluster Jobs}     

\author{
	{\bfseries Andresen. Dan$^1$, Hsu. William$^2$, Yang. Huichen$^2$, and Okanlawon. Adedolapo$^3$}\\
	Department of Computer Science, Kansas State University\\ Manhattan, Kansas, USA\\
}

\begin{document}

	\maketitle                        

	\begin{abstract}
		We address the problem of predicting whether sufficient memory and CPU resources have been requested for jobs at submission time.  For this purpose, we examine the task of training a supervised machine learning system to predict the outcome - whether the job will fail specifically due to insufficient resources - as a classification task.  Sufficiently high accuracy, precision, and recall at this task facilitates more anticipatory decision support applications in the domain of HPC resource allocation. Our preliminary results using a new test bed show that the probability of failed jobs is associated with information freely available at job submission time and may thus be usable by a learning system for user modeling that gives personalized feedback to users.
	\end{abstract}

	\vspace{1em}
	\noindent\textbf{Keywords:}
	{\small  HPC, machine learning, predictive analytics, decision support, user modeling} 
	

	\section{Introduction}
	This work presents a new machine learning-based approach as applied to high-performance computing (HPC).  In particular, we seek to train a supervised inductive learning system to predict when jobs submitted to a compute cluster may be subject to failure due to insufficient requested resources. There exist various open-source software packages which can help administrators to manage HPC systems, such as the Sun Grid Engine (SGE) [1] and Slurm [2]. These provide real-time monitoring platforms allowing both users and administrators to check job statuses.  However, there does not yet exist software that can help to fully automate the allocation of HPC resources or to anticipate resource needs reliably by generalizing over historical data, such as determining the number of processor cores and the amount of memory needed as a function of requests and outcomes on previous job submissions.  Machine learning (ML) applied towards decision support in estimating resource needs for an HPC task, or to predict resource usage, is the subject of several previous studies [3]--[7]. Our continuing work is based on a new predictive test bed developed using {\it Beocat}, the primary HPC platform at Kansas State University, and a data set compiled by the Department of Computer Science from the job submission and monitoring logs of {\it Beocat}.
	\\
	\indent Our purpose in this paper is to use supervised inductive learning over historical log files on large-scale compute clusters that contain a mixture of CPUs and GPUs.  We seek initially to develop a data model containing ground attributes of jobs that can help users or administrators to classify jobs into equivalence classes by likelihood of failure, based on aggregate demographic and historical profile information regarding the user who submitted each job.  Among these are estimating (1) the failure probability of a job at submission time, (2) the estimated resource utilization level given submission history.  We address the predictive task with the prospective goal of selecting helpful, personalized runtime or postmortem feedback to help the user make better cost-benefit tradeoffs.  Our preliminary results show that the probability of failed jobs is associated with information freely available at job submission time and may this be usable by a learning system for user modeling that gives personalized feedback to users.
	\\
	\indent {\it Beocat} is the primary HPC system of Kansas State University (KSU). When submitting jobs to the managed queues of {\it Beocat}, users need to specify their estimated running time and memory. The Beocat system then schedules jobs based on these job requirements, the availability of system resources, and job-dependent factors such as static properties of the executable. However, users cannot in general estimate the usage of their jobs with high accuracy, as it is challenging even for trained and experienced users to estimate how much time and memory a job requires. A central hypothesis of this work, based upon observation of users over the 20-year operating life of {\it Beocat} to date, is that estimation accuracy is correlated with user experience in particular use cases, such as the type of HPC codes and data they are working with. The risks of underestimation of resource requirements include wasting some of these resources: if a user submits a job which will fail during execution because this user underestimates resource needs at submission, the job will occupy some resources after submission until it has been identified as having failed by the queue management software of the compute cluster. This situation will not only affect the available resources but also affect other jobs in the cluster's queues. This is a pervasive issue in HPC, not specific to {\it Beocat}, and cannot by solved solely by proactive HPC management. We therefore address it by using supervised learning to build a model from historical logs.  This can help save some resources in HPC systems and also yield {\bf recommendations} to users such as estimated CPU/RAM usage of job, allowing them to submit more robust job specifications rather than waiting until jobs fail. 
	\\
	\indent In the remainder of this paper, we lay out the machine learning task and approach for {\it Beocat}, surveying algorithms and describing the development of a experimental test bed.
	
	\section{Machine Learning and Related Work}
	
	As related above, the core focus and novel contribution of this work are the application of supervised machine learning to resource allocation in HPC systems, particularly a predictive task defined on compute clusters. This remains an open problem as there is as yet no single machine learning representation and algorithm that can reliably help users to predict job memory requirements in an HPC system, as been noted by researchers at IBM [8]. However, it is still worth while to try to find some machine learning technologies that could help administrators better anticipate resource needs and help users make more cost-effective allocation choices in an HPC system. Different HPC system have different environments; our goal is to improve resource allocation in our HPC system. Our objectives go beyond simple CPU and memory usage prediction towards data mining and decision support. [9] There are thus two machine learning tasks on which we are focused: {\bf (1) regression} to predict usage of CPU and memory, and {\bf (2) classification} over job submission instances to predict job failure after the submission. We also try to train different models with different machine learning algorithms. The following part is the machine learning algorithms involved in our experiment.
	
	\subsection{Test bed for machine learning and decision support}
	
	A motivating goal of this work is to develop an open test bed for machine learning and decision support using {\it Beocat} data.  {\it Beocat} is at present the largest in the state of Kansas. It is also the central component of a regional compute cluster that provides a platform for academic HPC research, including many interdisciplinary and transdisciplinary users.  This is significant because user experience can vary greatly by both familiarity with computational methods in their domain of application and HPC platform.  Examples of application domain-specific computational methods include data integration and modeling, data transformations needed by the users on their own raw data, and algorithms used to solve their specific problems.  Meanwhile, the HPC platform depends design choices such as programming languages, scientific computing libraries, the parallel computing architecture (e.g., a parallel programming library versus MapReduce or ad hoc task parallelism), and load balancing and process migration methods, if any.
	
	{\bf Precursors: feature sets and ancillary estimation targets.} Attendant upon the development of a test bed are specific technical objectives in the form of actionable decision support tasks such as: ``Approximately what will it cost to migrate this job to a compute cloud such as Amazon Web Services, Azure, or Google Cloud, what will it cost to run on {\it Beocat}, and what are the the costs of various hybrid solutions?'' and ``Which of the following is expected to be most cost-effective based on these estimates and historical data?''  This in turn requires data preparation steps including integration and transformation.  
	
	{\bf Prediction targets and ground truth.} Data transformations on these logs allow us to define a basic regression task of predicting the CPU and memory usage of jobs, and the ancillary task of determining whether a job submission will fail due to this usage exceeding the allocation based on the resources requested by a user.  The ground truth for these tasks comes from historical data collected from {\it Beocat} over several years, which presents an open question of how to validate this ground truth across multiple HPC systems. This is a challenge beyond the scope of the present work but an important reason for having open access data: so that the potential for cross-system transfer can be assessed as a criterion, and cross-domain transfer learning methods can be evaluated.
	
	\subsection{Defining learning tasks: regression vs. classification}
	
	As explained above, this work concerns {\it learning to predict} for two questions: the numerical question of estimating the quantity of resources used (CPU cycles and RAM), and the yes-no question of whether a job will be killed.  The first question, CPU/RAM estimation, is by definition a discrete estimation task. However, in its most general form, the integer precision needed to obtain multiple decision support estimates such as those discussed in the previous section, and to then generate actionable recommendations for the available options, makes this in essence a continuous estimation task (i.e., regression). The section question is binary classification (i.e., concept learning, with the concept being "job killed due to resource underestimate").  
	
	Beyond the single job classification task, we are interested in the formulation of classification tasks for users - that is, assessing the level of expertise and experience. These may be independent factors as documented in the description of the test bed.
	
	\subsection{Ground Features and Relevance} 
	
	A key question is that of {\it relevance determination} - how to deal with increasing numbers of irrelevant features. [9]--[10]  In this data set, ground features are primitive attributes of the relational schema (and simple observable or computable variables). For our HPC predictive analytics task, this is initially a less salient situation, because the naive version of the task uses only {\bf per-job} features or predominantly such features, but becomes increasingly the case as {\bf per-user} features are introduced.

	\indent {\bf Linear regression} is a simple but powerful algorithm that we selected as a baseline for our prediction task. We chose to use multiple linear regression to simultaneously predict both CPU and memory usage for an HPC system. The dependent variables thus include CPU and memory usage, while the independent variables are the features presented in Table I. We can find a linear equation such as $ y = a_0 + a_1x_1 + a_2x_2 + \cdots + a_nx_n $ to fit the data set with minimum distance for each data point to the fitting line. There are various loss functions (i.e., ordinary least squares, ridge) to measure the fitness of linear regression models. By minimizing the loss functions, we can optimize our models. We use different regression and classification models in the experiment as follows.
	
	\indent {\bf Ordinary Least Squares}: this model fits a target function by minimizing the sum of squares of the difference between observations and values predicted by a linear approximation. The purpose of using this model is to minimize the error of estimations in order to optimize the linear regression model.
	
	\indent {\bf LassoLarsIC}: this linear model is trained with an L1 regularizer, which solves for the minimum least-squares penalty. These are favored in models with many zero weights (i.e., irrelevant features). Also, this model can be validated using the Akaike information criterion (AIC) or Bayes Information criterion (BIC).
	
	\indent {\bf ElasticNetCV}: this linear model is trained with weighted combination of L1 and L2 regularizer, and it can be used to set the parameters to determine which regularizer (L1, L2 or the combination between them) could be used by cross-validation.  
	
	\indent {\bf Ridge Regression}: this linear model is trained with an L2 regularizer, which can deal with overfitting issue. Compared with Ordinary Least Squares, Ridge Regression has a more stable feature by introducing a penalty to reduce the size of coefficients and avoid the overfitting problems that Ordinary Least Squares may have. 
	
	\indent {\bf CART: Classification and Regression Trees (CART)} is a non-linear model that we also selected for our experiment. This model mainly includes two types of decision trees: classification trees and regression trees. Classification trees have categorical variables and are used label the class of variables. Regression trees have continue variables and are used for predications.
	
	\indent {\bf Logistic Regression}: this model is a predictive analysis technique employed when the dependent variable is binary. Logistic regression is used to describe data and to explain the relationship between one dependent binary variable and one or more nominal, ordinal, interval or ratio-level independent variables.
	
	\indent {\bf Gaussian Naive Bayes}: this model is based on Naive Bayes with prior as Gaussian distribution. Gaussian Naive Bayes extends Naive Bayes to real-valued attributes by estimating the mean and the standard deviations generated by our data set. The representation for Gaussian Naive Bayes is to calculate the probabilities for input values for each class, and store the mean and standard deviations for them.
	
	\indent {\bf Random Forest Classification}: this model works by constructing a multitude of decision trees at training time and outputting the class that is the mode of the classes (classification) or mean prediction (regression) of the individual trees. It is a supervised learning algorithm based on an ensemble of decision trees as the learning representation.

	\section{Experiment Design}
	In this section, we describe the acquisition and preparation of training data for machine learning, the principles governing our feature extraction and selection process, and design choices for learning algorithms and their parameters.
	
	\subsection{Data Preparation and Feature Analysis}
	Our experiment is based on the SGE log file which is recorded and used by the Beocat system management software. The raw data set covers an 8-year history of job information which has been run on Beocat from 2009 to 2017. It has around twenty million instances, and has forty-five attributes for each instance. 
	\\
	\indent The purpose of the machine learning experiment is to train a regression model to predict CPU and memory usage, and to determine if submitted jobs will fail, based on requested resources and other {\em submission-time} attribute, those known after a job is submitted but before it is executed. At that time, a monitoring system can only provide very limited information, including: basic demographic information about the submitting user, such as their name, institution, department, etc., plus information available to the job scheduling system such as resource requests, particularly the estimated job running time and the expected maximum memory needed during runtime. However, the raw data set does not have additional forensic user such as the their job title (faculty, staff, student, or other), degree program for a student (graduate or undergraduate), home department, etc. We obtain forensic information about users, by using public services such as the Lightweight Directory Access Protocol (LDAP) [11] command on the HPC system, and use this to augment the raw data set.
	\\
	\indent Because only limited features can be used during job submission, we also consider the behavior of users, by analyzing features for user modeling [12]--[13] and using them for resource usage prediction by regression and job outcome prediction by classification. 
	
	\subsection{Feature Construction for User Modeling}
	A driving hypothesis of this work is that augmenting {\em per-job} features with {\em per-user} features can provide a {\bf persistent context of user expertise} that can improve the accuracy, precision, and recall of learned models. The rationale is that the skills and expertise level of a user are variables that: (1) change at a time scale much greater than job-by-job; (2) are based on low-level submission-time attributes; and (3) can be estimated more accurately as a user submits more jobs. These are tied to mid-range objectives of this research, namely, formulating testable statistical hypotheses regarding observable variables of the user model. For example, this work is motivated by the conjectures that: (1) the expertise level expressed as a job failure rate can be predicted over time; (2) per-user features can incrementally improve the precision, recall, and accuracy of the predictive models for resource usage and job failure achieved by training with submission-time per-job features only; and (3) the variance of per-user features tends to be reduced by collecting more historical data. 
	\\
	\indent Support for these hypotheses would indicate that individual user statistics known prior to job submission time, such as the job failure rate and margin of underestimates in resource requests for a user, can be cumulatively estimated. This would also pose further interesting questions of whether social models of users, from rudimentary "expertise clusters" to detectable communities of users with similar training needs and error profiles, can be formed by unsupervised learning. A long-term goal of this research is to identify such communities by applying network analysis (link mining or graph mining) algorithms to linked data such as property graphs of users and their jobs. [14]--[17]
	\\
	\indent The primitive per-user attributes that are computed are simply average usage statistics for CPU and memory across all jobs submitted by a user. We begin with these ground attributes because the defined performance element for this research is based on the initial machine learning task of training regression and classification models to predict per-job usage of CPU and memory. Subsequent modeling objectives, such as forming a {\em causal explanation} for this prediction target, also depend on capturing potential influents of job failure such as user inexperience, and imputing a quantifiable and relevant experience level based on user self-efficacy and track record (i.e., per-user attributes as a time series).\\
	\indent We chose the average of CPU usage, average of memory usage, average of running time requested, and average of memory requested for each user as our per-user behavioral features. The data transformations used to preprocess the raw data (consisting of one tuple per job) included aggregation (roll-up) across jobs. This groups across jobs, by user, computing the average usage value for CPU, memory, requested time and requested memory from raw data set, and projects the resulting columns: 
	
	\begin{center}
		{\bf $ \mathcal{G}_{average(CPU, memory, reqTime, reqMem)}(User) $ }
	\end{center}
	
	We then re-joined the per-user aggregates (rolled-up values resulting from the above grouping operation across jobs) into the raw data set:
	\begin{center}
		{\bf $ New\:data := per-user\:aggregates \bowtie_{user} raw\:dataset $ }
	\end{center}
	
	This results in one row per job again, with values from each per-user relation replicated across all jobs submitted by that user.
	
	\indent The raw data set stretches over 8 years, and includes some invalid values before cleaning, such as data with missing values. We restricted the data set to the past three years to enforce a higher standard of data quality. This more recent historical data consists of 16 million instances and admits a schema of forty-five raw attributes. To produce a representative data set that can be used to train models using machine learning libraries on personal computers, we selected one million instances from this data set uniformly at random. Because average number of submitted jobs per person is more than two thousand, we filtered out the users who submitted fewer than two hundred jobs. The schema was also reduced to eliminate some raw attributes that are known to be redundant, correlated, or ontologically irrelevant.  This resulted in a data set with one million instances and 18 selected features, which is described in Table 1.
	
	\begin{table}[htb]\centering
		\caption{Feature Selected}\label{t_sim}
		\begin{tabular}{@{}lcl@{}} \toprule		
			Feature & Type & Describe \\ \midrule
			failed& Numeric& Indicate job failed or not, 0/1\\
			cpu& Numeric& CPU usage (predicted variable)\\
			maxvmem&Numeric& Memory usage (predicted variable)\\
			id&	Numeric	&User id\\
			reqMem&	Numeric	&Memory requested at job submission\\
			reqTime	&Numeric&Time requested at job submission\\
			project&Numeric&Project which was assigned to job\\
			aCPU&Aggregate&Average CPU usage for user\\
			aMaxmem&Aggregate&Average memory usage for user\\
			aReqtime&Aggregate&Average running time requested by user\\
			aReqmem&Aggregate&Average of memory requested by user\\
			p\_Faculty&Categorical&Role of user\\
			p\_Graduate&Categorical&Role of user\\		
			p\_PostDoc&Categorical&Role of user\\
			p\_ResearchAss&Categorical&Role of user\\
			p\_Staff&Categorical&Role of user\\
			p\_UnderGra&Categorical&Role of user\\
			p\_Unknowing&Categorical&Role of user\\
			\bottomrule
		\end{tabular}
	\end{table} 
	
	\indent Entries of type "[Numeric] Aggregate" in Table 1 constitute all (and only) those that are per-user behavioral ground features. 
	
	\subsection{Machine Learning Implementation}
	To handle the various types of data, we did standardization for the data set before training the prediction model. We used the {\tt scikit-learn} [18]--[19] Python library for our experiment implementation.
	
	\subsection*{Prediction Techniques}
	
	\begin{itemize}
		\item Linear Regression: Linear Regression: we set up all parameters by default, such as normalize = False, n\_jobs = 1.
		\item LassoLarsIC Regression: LassoLarsIC Regression: we choose `aic (Akaike information criterion)' as the criterion's parameter which is used to identify the goodness of fit.
		\item ElastcNetCV Regression: ElasticNetCV Regression: l1\_ration parameter in this model means which regularization (L1 or L2) you want to use for training your model. We choose 0.5 (penalty is a combination of L1 and L2) which is a default value for this parameter, as changing to other values for this parameter does not help to improve the result based on our data set. We choose default value `None' for the parameter of alpha.
		\item Ridge Regression: Ridge Regression: in this model, alpha represents the strength of regularization, we choose 0.5 which is the default value for this parameter. `auto' is set up for the parameter of the solver, which indicates that the model would choose the solver automatically based on the type of data set, such as svd, cholesk.
		\item CART Regression: we use `mse (mean squared error)' as the criterion parameter, and `best' instead of `random' as the parameter of the splitter. We use `None' for the parameter of the max\_depth, because this option will not affect the model result in our data set.
	\end{itemize}
	
	\subsection*{Classification Techniques}
	
	\begin{itemize}
		\item Logistic Regression: Logistic Regression: we choose l2 regularization - `l2' as penalty parameter and `liblinear' as solver parameter which refers to the optimization method for finding the optimum of the objective function.
		\item CART Classification: CART Classification: we choose `gini' as the criterion parameter rather than `entropy'. The `best' is used as splitting parameter. The parameter of maximum depth of the tree max\textunderscore depth is set up as `None'.
		\item GaussianNB Classification: GaussianNB: there is only one parameter `priors' and it is used by default value `None'. This parameter means the prior probabilities of the classes.
		\item Random Forest Classification: Random Forest Classification: `gini' is chosen to be the criterion parameter rather than `entropy'. `None' is set up for max\textunderscore depth parameter.
	\end{itemize}
	
	\section{Evaluation}
	In this section, we describe the evaluation strategy we used for the experiment, and present results consisting of quantitative metrics, followed by their qualitative interpretation in the context of the prediction tasks defined in the preceding experiment design section.
	\\
	\indent The applicative phase generates predicted CPU and memory usage from new instances (unlabeled examples). The long-term performance of a user is taken into consideration via per-user features given as input to the regression model, such as average CPU usage across all jobs submitted by a user, rather than past jobs submitted by a user up to the present job. That is, per-user features are replicated rather than cumulatively calculated. We computed per-user features by aggregating (rolling up) values across tuples (one row per job) to obtain one row per user, for each new columns (one column per aggregation operator).  In general, different aggregation operations such as {\tt AVERAGE}, {\tt MIN}, {\tt MAX}, and {\tt COUNT} can be used; in our experiments reported in this work, {\tt AVERAGE} is the only aggregate value calculated.
	\\
	\indent The changes in accuracy and F1 from the baseline data set to this new training data set are computed, to assess the incremental gain of adding per-user aggregate features. For predicting CPU usage, these aggregates consist of the average CPU usage (aCpu) and average requested time (aReqtime), computed across previously submitted jobs for each user to produce the CPU training data. Similarly, for memory usage prediction, the average memory usage (aMaxmem) and average requested memory (aReqmem) across previous submitted jobs for a user, are computed to produce training data. We use the R-squared statistic, a common measure of accuracy, to quantitatively evaluate our regression models. The R-squared values of different models are compared. Table 2 and 3 describe and list results for different machine learning algorithms applied to the version of the data set with per-user aggregate features (marked "True" in the second column).

	The semantics of the data model in Table 2 and 3 are as follows: 
	
	\begin{itemize}
		\item LLIC: LassoLarsIC Regression
		\item ENCV: ElasticNetCV Regression
		\item Ridge: Ridge Regression
		\item CART: CART Regression
		\item Aggregate Features (Table 2): aCPU + aReqtime
		\item Aggregate Feature (Table 3): aMaxmem + aReqMem
	\end{itemize}
	
	\begin{table}[htb]\centering
		\caption{CPU usage prediction with Regression}\label{t_sim}
		\begin{tabular}{@{}lccc@{}} \toprule
			Model & Per-User Features & R squared (\%) & Time (second) \\ \midrule 
			
			LinearRegression & True & 15.86 & 0.448 \\ 
			
			LinearRegression & False & 6.01 & 0.343 \\
			
			LLIC & True & 15.85 & 0.445 \\
			
			LLIC & False & 6.01 & 0.398 \\
			
			ENCV & True & 14.99 & 6.679 \\ 
			
			ENCV & False & 6.01 & 6.381 \\ 
			
			Ridge  & True & 15.86 & 0.224 \\
			
			Ridge  & False & 6.01 & 0.211 \\
			
			CART & True & 27.86 & 3.090 \\
			
			CART & False & 29.90 & 2.205 \\
			\bottomrule
		\end{tabular}
	\end{table}

	\begin{table}[htb]\centering
		\caption{Memory usage prediction with Regression}\label{t_sim}
		\begin{tabular}{@{}lccc@{}} \toprule 
			Model & Per-User Features & R squared (\%) & Time (second) \\ \midrule 
			
			LinearRegression & True & 23.11 & 0.406 \\ 
			
			LinearRegression & False & 16.70 & 0.348 \\
			
			LLIC & True & 23.11 & 0.451 \\
			
			LLIC & False & 16.70 & 0.410 \\
			
			ENCV & True & 23.11 & 6.387 \\ 
			
			ENCV & False & 16.70 & 7.273 \\ 
			
			Ridge  & True & 23.11 & 0.249 \\
			
			Ridge  & False & 16.70 & 0.202 \\
			
			CART & True & -23.23 & 2.108 \\
			
			CART & False & -27.12 & 1.472 \\
			\bottomrule
		\end{tabular}
	\end{table}
	
	\indent For the classification task, per-user features are taken into account. We re-joined the per-user, across-job numeric aggregates (in Table 1) in classification task which is different from the regression tasks. The accuracy score which is provided by {\tt scikit-learn} [18] has been used in classification model measurement. The accuracy score represents the ratio of samples which has been classified to the total samples in our classification model. A higher percentage of accuracy score implies a more accurate classification result in the classification model. F1 score, also known as {\it F-measure}, is the harmonic mean of the precision and recall of our model. This score is a weighted average of the precision and recall of a model, ranging from 0 (worst model) to 1 (best model). In addition, F1 score has the equivalent indication for both precision and recall. Table 4 shows the results of various measurements for different classification models.
	
	\begin{table}[htb]\centering
		\caption{Classification results}\label{t_sim}
		\begin{tabular}{@{}lcccc@{}} \toprule  
			Model & Per-User Features & Accuracy(\%) & F1(\%) & Time(second) \\ \midrule 
			
			LR & True & 95.20 & 93 & 245.642 \\ 
			
			LR & False & 95.22 & 93 & 71.680 \\
			
			CART & True & 96.87 & 96 & 30.910 \\
			
			CART & False & 96.87 & 96 & 19.642 \\
			
			GNB & True & 92.43 & 92 & 8.465 \\ 
			
			GNB & False & 92.00 & 92 & 6.107 \\ 
			
			RF  & True & 96.87 & 96 & 64.356 \\
			
			RF  & False & 98.86 & 96 & 59.743 \\
			\bottomrule
		\end{tabular}
	\end{table}
	
	The legend for Table 4 is as follows:
	
	\begin{itemize}
		\item LR: Logistic Regression
		\item CART: CART Classification
		\item NB: Gaussian Naive Bayes Classification
		\item RF: Random Forest Classification
	\end{itemize}

	As Tables 2 and 3 indicate, there is a substantial gain in (the relatively low positive values of) R-squared, the coefficient of determination, as a result of using per-user aggregates, and that this is borne out across different regression models. However, for the classification models trained in this work, these gains are not matched by a commensurate gain in accuracy, precision, or recall.  The marginal increase in R-squared suggests that the approach of joining per-job tuples with replicated per-user statistics is still promising.
	
	\section{Conclusions and Future Work}
	
	In this section, we conclude with a brief discussion of findings, immediate priorities for continuing work, and next steps for the overall approach presented in this work.
	
	\subsection{Summary}
	
	In this work we have developed the initial version of a test bed for machine learning for predictive analytics and decision support on a compute cluster.  As derived, our regression-based usage prediction and classification tasks are disparate and require different machine learning models.  Our experiment results show that where regression can be used for the prediction task and regression-based classification for the overall concept of a job being killed due to resource underestimates, the learning {\bf algorithms} also differ in most cases.  
	
	\subsection{Findings and Conclusions}
	
	The results demonstrate some potential for learnability of the regression and classification target functions using the historical data in our initial test bed. We can use CART regression to predict CPU usage as its data set is more like a non-linearity distribution, but CART regression does not satisfy memory usage prediction at all, because the data set of memory is distributed nearly linearly. For the classification task, all the models got high accuracy score. Thus, we should consider the model that is the least time-consuming, which is Gaussian Naive Bayes in our experiment.
	
	Our experiment brings a possible solution for applying machine learning to the HPC system, but we still need to focus on improving the prediction accuracy for CPU and memory usage. To predict CPU and memory usage during the job submission is still a difficult task due to the insufficient features that can be used, and there are no conspicuous relationships between CPU and memory usage and the existing features from the data set that could be captured for prediction.  
	
	Incremental gain in R-squared is observed for per-user aggregate features, but no appreciable gain in accuracy or F1 is observed for any inducers (models for supervised inductive learning) that were used. Thus, the experiment is inconclusive as to measurable impact, but further experimentation with other causal models of user behavior such as Bayesian networks and other graphical models of probability.  
	
	\subsection{Current and Future Work}
	
	Some promising future directions suggested by the above results and interpretation include exploring the user modeling aspect of predictive analytics for HPC systems.  We are developing a survey instrument to collect user self-efficacy and background information about training.  This can be used to segment users in a data-driven way: based on clustering algorithms rather than merely stratifying them by years of experience, even by category. This would also facilitate exploration of the hypotheses outlined in \textsection 3.2 regarding change in imputed user expertise.
	
	Moreover, given the availability of user demographic data, there appears to be potential to incorporate a social aspect of the data to broaden the scope and applicability of the test bed - by learning from linked data and building network models including but not limited to detecting "communities of expertise" within the set of users on an HPC system. [14]--[17]
	
	\begin{figure}[h]
		\caption{Types of links in an example social network. [17]} 
		\includegraphics[width=\linewidth]{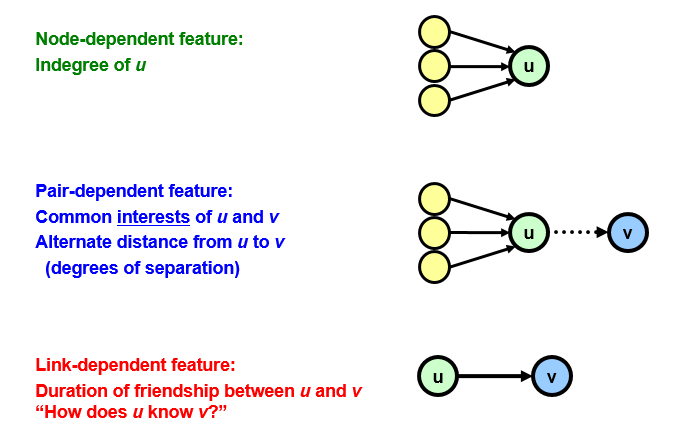}
	\end{figure}
	
	The existence of a relationship existing between two users in a social network can be identified by an inference process or by simple classification.  Although the inference steps may be probabilistic, logical, or both, the links themselves tend to be categorical.  As Figure 1 shows, they can be dependent purely on single nodes, local topology, or exogenous information. [17]  In addition to using the structure of the known graph, common features of candidates for link existence (in this domain, profile similarity; in others, friendship, trust, or mutual community membership) include similarity measures such as the number of job failures of the same type, or some semantically-weighted measure of similarity over job outcome events.
	
	\section{Acknowledgements}
	
	This work was also originally supported in part by the U.S. National Science Foundation (NSF) under grant CNS-MRI-1429316. The authors thank Dave Turner, postdoctoral research associate, and Kyle Hutson, a system administrators of Beocat, for consultation on the original SGE and Slurm log file formats. The authors also thank all of the students of CIS 732 {\em Machine Learning and Pattern Recognition} in spring, 2017 and spring, 2018 who worked on term projects related to the prediction tasks presented in this paper, including: Mohammedhossein Amini, Kaiming Bi, Poojitha Bikki, Yuying Chen, Shubh Chopra, Sandeep Dasari, Pruthvidharreddy Dhodda, Sravani Donepudi, Cortez Gray, Sneha Gullapalli, Nithin Kakkireni, Blake Knedler, Atef Khan, Ryan Kruse, Mahesh Reddy Mandadi, Tracy Marshall, Reza Mazloom, Akash Rathore, Debarshi Saha, Maitreyi Tata, Thaddeus Tuck, Sharmila Vegesana, Sindhu Velumula, Vijay Kumar Venkatamuniyappa, Nitesh Verma, Chaoxin Wang, and Jingyi Zhou.


\begin{thebibliography}{1}
		\providecommand{\url}[1]{#1}
		\csname url@rmstyle\endcsname
		\providecommand{\newblock}{\relax}
		\providecommand{\bibinfo}[2]{#2}
		\providecommand\BIBentrySTDinterwordspacing{\spaceskip=0pt\relax}
		\providecommand\BIBentryALTinterwordstretchfactor{4}
		\providecommand\BIBentryALTinterwordspacing{\spaceskip=\fontdimen2\font plus
			\BIBentryALTinterwordstretchfactor\fontdimen3\font minus
			\fontdimen4\font\relax}
		\providecommand\BIBforeignlanguage[2]{{%
				\expandafter\ifx\csname l@#1\endcsname\relax
				\typeout{** WARNING: IEEEtran.bst: No hyphenation pattern has been}%
				\typeout{** loaded for the language `#1'. Using the pattern for}%
				\typeout{** the default language instead.}%
				\else
				\language=\csname l@#1\endcsname
				\fi
				#2}}
		
		\bibitem{IEEE:confwithpaper}
		Wolfgang. Gentzsch, ``Sun grid engine: Towards creating a compute power grid,'' 
		in \emph{Cluster Computing and the Grid, 2001. Proceedings. First IEEE/ACM International Symposium on}, pp. 35-36. IEEE, 2001.
		
		\bibitem{IEEE:confwithpaper}
		Yoo, Andy B., Morris A. Jette, and Mark Grondona. ``Slurm: Simple linux utility for resource management,'' in \emph{Workshop on Job Scheduling Strategies for Parallel Processing}, pp. 44-60. Springer, Berlin, Heidelberg, 2003.
		
		\bibitem{IEEE: confwithpaper}
		Matsunaga, Andr{\'e}a and Fortes, Jos{\'e} AB. ``On the use of machine learning to predict the time and resources consumed by applications,'' in \emph{Proceedings of the 2010 10th IEEE/ACM International Conference on Cluster, Cloud and Grid Computing}, pp. 495-504. IEEE Computer Society, 2010.
		
		\bibitem{IEEE: journal}
		Bugbee, B., and Phillips, C., Egan, H., and Elmore, R., Gruchalla, K., and Purkayastha, A. "Prediction and characterization of application power use in a high-performance computing environment.'' \emph{Statistical Analysis and Data Mining: The ASA Data Science Journal} vol. 10, pp. 155-165, 2017.
		
		\bibitem{IEEE: confwithpaper}
		Berral, Josep Ll and Goiri, {\'I}{\~n}igo and Nou, Ram{\'o}n and Juli{\`a}, Ferran and Guitart, Jordi and Gavald{\`a}, Ricard and Torres, Jordi. ``Towards energy-aware scheduling in data centers using machine learning,'' in \emph{Proceedings of the 1st International Conference on energy-Efficient Computing and Networking}, pp. 215-224. ACM, 2010.
		
		\bibitem{IEEE: journal}
		Chou, J. S., Chiu, C. K., Farfoura, M., and Al-Taharwa, ``Optimizing the prediction accuracy of concrete compressive strength based on a comparison of data-mining techniques,''  \emph{Journal of Computing in Civil Engineering}., vol. 25, pp. 242-253, 2000.
		
		\bibitem{IEEE: confwithpaper}
		Li, J., Ma, X., Singh, K., Schulz, M., de Supinski, B. R., and McKee, S. A. ``Machine learning based online performance prediction for runtime parallelization and task scheduling,'' in \emph{Performance Analysis of Systems and Software, 2009. ISPASS 2009. IEEE International Symposium on}, pp. 89-100. IEEE, 2009.
		
		\bibitem{IEEEE: confwithpaper}
		Rodrigues, E. R., Cunha, R. L., Netto, M. A., and Spriggs, M. ``Helping HPC users specify job memory requirements via machine learning,'' in \emph{Proceedings of the Third International Workshop on HPC User Support Tools}, pp. 6-13, IEEE Press, 2016.
		
		\bibitem{IEEE: journal}
		Kohavi, R., and John, G. H., ``Wrappers for Feature Subset Selection'', \emph{Artificial Intelligence}, vol. 97, no. 1, pp. 273-324, 1997.
		
		\bibitem{IEEE: journal}
		Hsu, W. H., Welge, M., Redman, T., and Clutter, D.  ``High-Performance Commercial Data Mining: A Multistrategy Machine Learning Application.'' \emph{Data Mining and Knowledge Discovery}, vol. 6, no. 4, pp. 361-391, 2002.
		
		\bibitem{IEEE: book}
		Howes, T., and Smith, M, \emph{LDAP: programming directory-enabled applications with lightweight directory access protocol}. Sams Publishing, 1997.	
		
		\bibitem{IEEE: journal}
		Webb, G. I., Pazzani, M. J., and Billsus, D, ``Machine learning for user modeling,'' \emph{User modeling and user-adapted interaction}, vol. 11, pp. 19-29, 2001.
		
		\bibitem{IEEE: conference}
		Yao, Y., Zhao, Y., Wang, J., and Han, S, ``A model of machine learning based on user preference of attributes''. \emph{International Conference on Rough Sets and Current Trends in Computing}, pp. 587-596. Springer, Berlin, Heidelberg, 2006.
		
		\bibitem{IEEE: chapter}
		Yang, M., Hsu, W. H., and Kallumadi, S. ``Predictive Analytics of Social Networks: A Survey of Tasks and Techniques.'' In Hsu, W. H., ed., \emph{Emerging Methods in Predictive Analytics: Risk Management and Decision-Making}, pp. 297-333. IGI Global, 2016.
		
		\bibitem{IEEE: tutorial}
		Beutel, A., Akoglu, L., and Faloutsos, C. ``Graph-Based User Behavior Modeling: From Prediction to Fraud Detection''. In Cao, L., Zhang, C., Joachims, T., Webb, G. I., Margineantu, D. D., Williams, G., eds. \emph{Proceedings of the 21th ACM SIGKDD International Conference on Knowledge Discovery and Data Mining (KDD 2015), Tutorial Program}, pp. 2309-2310, 2015.
		
		\bibitem{IEEE: journal}
		Adedoyin-Olowe, M., Gaber, M. M., and Stahl, F., ``A Survey of Data Mining Techniques for Social Media Analysis'', vol. 2014, pp. 1-25, 2014.
		
		\bibitem{IEEE: conference}
		Hsu, W. H., Lancaster, J. P., Paradesi, M. S. R., and Weninger, T. ``Structural Link Analysis from User Profiles and Friends Networks: A Feature Construction Approach'', \emph{Proceedings of the 1st International Conference on Weblogs and Social Media (ICWSM 2007)}, pp. 75-80, 2007.
		
		\bibitem{IEEE: journal}
		Pedregosa, F., Varoquaux, G., Gramfort, A., Michel, V., Thirion, B., Grisel, O., Blondel, M., Prettenhofer, P., Weiss, R., Dubourg, V., Vanderplas, J., Passos, A., Cournapeau, D., Brucher, M., Perrot, M., and Duchesnay, E. ``Scikit-learn: Machine Learning in Python'' \emph{Journal of Machine Learning Research}, vol. 12, pp. 2825-2830, 2011.
		
		\bibitem{IEEE: book}
		Massaron, L., and Boschetti, A, \emph{Regression Analysis with Python}. Packt Publishing Ltd, 2016.
		
	\end{thebibliography}
\end{document}